\documentclass[referee]{raa}            

\usepackage{graphicx,times}             

\begin{document}

   \title{V470 Cas and GSC 2901-00089, Two New Double-mode Cepheids
   }

   \volnopage{Vol.0 (200x) No.0, 000--000}      
   \setcounter{page}{1}          

   \author{A. V. Khruslov
      \inst{1,2}
   \and A.V. Kusakin
      \inst{3}
   }

   \institute{Sternberg Astronomical Institute, Moscow University,
             Moscow 119992, Russia; {\it khruslov@bk.ru }\\
        \and
             Institute of Astronomy, Russian Academy of Sciences, Moscow, 119017, Russia \\
        \and
             Fesenkov Astrophysical Institute, Almaty, 050020, Kazakhstan; {\it un7gbd@gmail.com}\\
   }

   \date{Received~~2009 month day; accepted~~2009~~month day}

\abstract{ We present a photometric study of two new double-mode
Cepheids, pulsating in the first and second overtones modes: V470
Cas and GSC 2901-00089. For the search of the double-mode
variability, we used all available observations from the
ROTSE-I/NSVS and SuperWASP online public archives. Our multicolour
CCD observations in the $B$, $V$ and $R$ bands in Johnson's system
confirm the double periodicity of these variables. We study period
variations of the two stars; variations of the first overtone
periods were reliably detected. In addition, we consider the
Petersen diagram for the Galactic 1O/2O Cepheids.
\keywords{techniques: photometric --- stars: variables: Cepheids
--- stars: individual: V470 Cas, GSC 2901-00089} }

   \authorrunning{A. V. Khruslov \& A.V. Kusakin }            
   \titlerunning{V470 Cas and GSC 2901-00089, Two New Double-mode Cepheids}  

   \maketitle

%
%
\section{Introduction}           
\label{sect:intro}

Classical Cepheid are radially pulsating periodic variable stars.
Pulsations of single-period Cepheids occur in the fundamental mode
(F) or in the first overtone mode (1O). In addition, several
examples of possible second overtone (2O) Cepheids are known. The
double-mode Cepheids (or beat Cepheids) pulsate simultaneously in
two radial pulsation modes: in the fundamental mode and in the
first overtone mode (F/1O) or in the first and second overtone
mode (1O/2O). The period ratio is a good indicator of the excited
modes. Characteristic period ratios are $0.69 < P_1/P_0 < 0.73$
and $P_2/P_1 = 0.80$. The mass of the Cepheid can be derived using
the two pulsation periods only. The first results based on the
linear theory of stellar pulsations were obtained by Petersen
(\cite{pete73}). For recent results of non-linear modelling of
double-mode Cepheids, see Smolec \& Moskalik (\cite{smol10}).

Double-mode Cepheids are known for more than half a century.
Oosterhoff (\cite{oost57}) was the first to detect a group of
$\delta$~Cephei variable stars with a large scatter in their
photoelectric phased light curves, because of superposition of two
oscillations. Later, double periodicity oscillations in the
fundamental mode and first overtone were confirmed for most of
these stars. To date, a little more than 30 F/1O Cepheids are
known.
\newpage

The first Cepheid pulsating in the first and second overtones was
discovered by  Mantegazza (\cite{mant83}). It was CO~Aur, which
long remained the only 1O/2O Cepheid. The next variables of this
type in our Galaxy were V1048 Cen (Beltrame \&
Poretti~\cite{belt02}), V767 Sgr and V363 Cas (Hajdu et al.
~\cite{hajd09}).

In 2009--2012, 12 new 1O/2O Cepheids were detected by one of the
authors (Khruslov ~\cite{khru09a},~\cite{khru09b},
~\cite{khru10a},~\cite{khru10b}, ~\cite{khru12}) from data of the
ASAS-3 and NSVS surveys. Of these, three short-period variables
(GSC 6567-01616, TYC 0717 01091 1, and GSC 6558-01290) were
originally classified as double-mode RR Lyrae stars. Later, we
changed the classification of these three stars due to their small
galactic latitude (respectively $b=+3^{\circ}.0$, $-6^{\circ}.2$,
and $+2^{\circ}.0$). 1O/2O RR Lyrae stars at large distances from
the Galactic plane are not known, so we consider these stars along
with 1O/2O Cepheids. Other examples of such stars are V1719 Cyg,
considered an RRC variable in the GCVS according to Mantegazza \&
Poretti (\cite{mant86}) ($P_1=0.2673$ days, $P_2/P_1 = 0.7998$,
$b=+2^{\circ}.6$), and V798 Cyg, a double-mode HADS star according
to Musazzi et al. (\cite{musa98}) ($P_1 = 0.1948$ days,
$P_2/P_1=0.8010$, $b=+4^{\circ}.6$).

The OGLE-III project discovered seven more 1O/2O Cepheids near the
galactic center (Soszynski et al.~\cite{sosz11a},~\cite{sosz11b};
Pietrukowicz et al.~\cite{piet13}); three of them also have very
short periods in the $0^d.24 - 0^d.29$ range. In addition,
Khruslov (\cite{khru13}) reported a possible first and second
overtone double periodicity of MS~Mus and TYC 8308 02055 1.

According to the OGLE-III survey, double-mode Cepheids are well
represented in the Large (LMC) and Small (SMC) Magellanic Clouds:
in the LMC, there are 61 F/1O and 203 1O/2O Cepheids; in the SMC,
there are 59 F/1O and 215 1O/2O Cepheids (Soszynski et
al.~\cite{sosz08b},~\cite{sosz10}).

Also, two triple-mode stars with periods in the Cepheid range are
known in the Galaxy: AC~And (Florya~\cite{flor37}, Fitch \&
Szeidl~\cite{fitc76}) and V823~Cas (Antipin~\cite{anti97}); they
show F/1O/2O pulsations. In the Magellanic Clouds, several
triple-mode Cepheids are known (Soszynski et
al.~\cite{sosz08a},~\cite{sosz10}).

In this paper, we present a photometric study of two new
double-mode Cepheids, pulsating in the first and second overtone
modes: V470 Cas and GSC 2901-00089.


\section{Observations and data reduction}
\label{sect:Obs}

To search for double-mode variability, we used all available
observations from the
ROTSE-I/NSVS (Wozniak et
al.~\cite{wozn04}, http://skydot.lanl.gov/nsvs) and
SuperWASP (Butters et al.~\cite{butt10}, http://wasp.cerit-sc.cz/form) 
online public archives. The SuperWASP
observations are available as FITS tables, which were converted
into ASCII tables using the OMC2ASCII program as described by
Sokolovsky (\cite{soko07}); we also used the SuperWASP FITS to
ASCII lightcurve conversion
service (http://scan.sai.msu.ru/swasp\_converter/). We
reliably classified GSC 2901--00089 as a 1O/2O Cepheid and
reported preliminary results in Khruslov (\cite{khru13}). Possible
double-mode variability of V470~Cas was suspected by us using
ROTSE-I/NSVS and SuperWASP data.
\newpage

To confirm the double-mode variability of these stars, we started
multicolor CCD observations in 2013. Our CCD observations in the
Johnson $B$, $V$ and $R$ bands were performed at the Tien Shan
Astronomical Observatory of the V.G.~Fesenkov Astrophysical
Institute, at the altitude of 2750~m above the sea level. The
observatory has two Zeiss 1000-mm telescopes. Most of our
observations were performed with the eastern Zeiss 1000-mm
reflector (the focal length of the system was $f=13380$~mm before
JD 2456500 and 6650~mm after this date; the detector was an Apogee
U9000 D9 CCD camera; the chip was cooled to $-40^{\circ}$~C). The
time interval of the observations for GSC 2901--00089 is JD
2456364--2456963 (March 12, 2013 -- November 1, 2014); for
V470~Cas, it is JD 2456575--2456964 (October~9, 2013 --
November~2, 2014). Additionally, for observations of V470~Cas
during two nights (JD 2456899 and 2456959), we used the newly
introduced western Zeiss 1000-mm reflector (the focal length of
the system was $f=13250$ mm, the detector being an Apogee F16M CCD
camera); during one night, JD 2456584, we used the 360-mm
Ritchey--Chretien telescope designed by V.B.~Sekirov (the focal
length of the system is 1440~mm; the detector was an ST-402 SBIG
CCD camera; the chip was cooled to $-20^{\circ}$~C).

Reductions were performed using the MaxIm DL aperture photometry
package. For GSC 2901--00089, we obtained a homogeneous
observations set. For V470~Cas, exposures of different lengths
were used, and we obtained a non-homogeneous observations set. In
addition, the small amplitude of the second oscillation, 2O,
requires minimal observation errors. Therefore, we averaged
individual values over time intervals of nearly the same duration,
each point being an average of 3--6 individual observations.

Information on the comparison stars and check stars for the two
Cepheids, used in our CCD photometry, is presented in Table 1.
Magnitudes of the comparison stars (in Johnson's $B$ and $V$
bands) were taken from the AAVSO Photometric All-Sky Survey
(APASS, http://www.aavso.org/download-apass-data)
catalog. The $R$-band observations could be presented only as
magnitude differences with respect to the comparison star. For GSC
2901--00089, the magnitudes differences $\Delta R$  in the $R$
band are $\Delta R=m_{\rm var}-m_{\rm comp}$; for V470 Cas,
$\Delta R=m_{\rm var}-m_{\rm comp}+1^m.916$.

The finding charts of the two Cepheids are displayed in Fig.~1.

We analyzed the time series using Deeming's method
(Deeming~\cite{deem75}) implemented in the WinEfk code written by
V.P. Goranskij.

\begin{table}
\begin{center}
\caption[]{ Comparison and check stars.}\label{Tab:publ-works}

 \begin{tabular}{llcc}
  \hline\noalign{\smallskip}
Variable  &   & V470 Cas  & GSC 2901--00089                \\
  \hline\noalign{\smallskip}
Comparison star  & Name  & GSC 3678--00722  & GSC 2901--00493                \\
  & Coordinates, J2000.0  & 01$^h$32$^m$12$^s$.75 +56$^{\circ}$30$^{\prime}$40$^{\prime\prime}$.0 & 04$^h$45$^m$06$^s$.34 +42$^{\circ}$57$^{\prime}$50$^{\prime\prime}$.9                \\
  & $V$ mag  & 13.937  & 12.827                \\
  & $B$ mag  & 14.755  & 13.326                \\
  \noalign{\smallskip}\hline
Check star  & Name  & GSC 3678--01408  & USNO-B1.0 1329--0132855                \\
  & Coordinates, J2000.0  & 01$^h$31$^m$51$^s$.71 +56$^{\circ}$27$^{\prime}$53$^{\prime\prime}$.6  & 04$^h$44$^m$59$^s$.64 +42$^{\circ}$57$^{\prime}$25$^{\prime\prime}$.5                \\
  \noalign{\smallskip}\hline
\end{tabular}
\end{center}
\end{table}

\newpage

\begin{figure}
   \centering
   \includegraphics[width=\textwidth, angle=0]{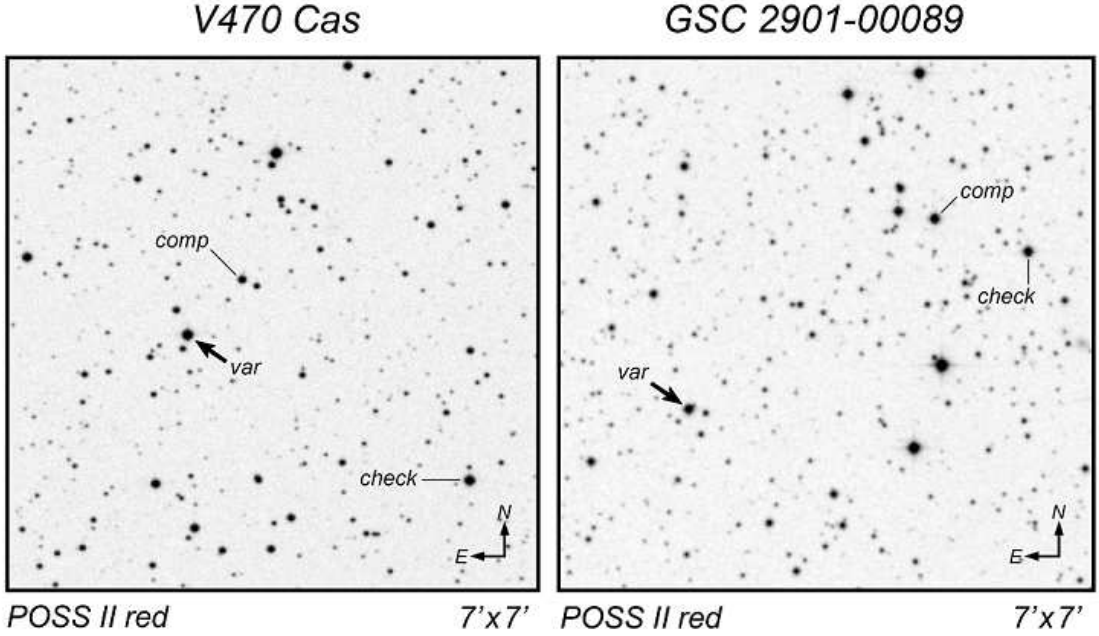}
   \caption{The finding charts.}
   \label{Fig1}
   \end{figure}

\section{Results}
\label{sect:data}

Our two program variable stars are new double-mode Cepheids,
pulsating in the first and second overtone modes. We studied
brightness variations of these stars using NSVS and SWASP data and
our CCD observations, identified periods of 1O and 2O pulsations,
and found period variations.

\subsection{V470 Cas}
\subsubsection{Earlier studies}

The variability of V470 Cas = S8459 ($\alpha=01^h32^m18^s.16,\
\delta = +56^{\circ} 29^{\prime} 58^{\prime\prime}.0$, J2000) was
discovered by Hoffmeister (\cite{hoff64}). The variable was
classified as a short-periodic variable (possibly eclipsing), the
variability range was $12^m.5 - 13^m.0$. The first study of the
variability of V470 Cas was published by Meinunger (\cite{mein68})
who classified the star as an eclipsing variable with the light
elements:

MinI~$(\rm {JD}) = 2429231.369 + 0^d.444692 \times E$.

The variability range is $13^m.0 - 13^m.5$.

Gessner \& Meinunger (\cite{gess73}) confirmed these light
elements but remarked on their being not quite certain. Eight
times of light minima were reported. The variable was included in
General Catalog of Variable Stars (Samus et al.~\cite{samu15})
based on this publication.

\newpage

Agerer et al. (\cite{ager96}) performed CCD-observations of
V470~Cas and studied plates of the Sonneberg Sky Patrol. It was
found that the star was not an eclipsing variable. V470~Cas is a
possible RR~Lyrae variable star with a long period and small
amplitude. The amplitude of variability in the instrumental system
(without filters) is $0^m.35$. Asymmetry of the light curve is
more typical for classical Cepheids ($M-m = 0^P.35$). During the
interval of observations (photographic observations: JD 2436200 --
2448862, CCD observations: JD 2449170 -- 2450013), the period of
variability changed. Therefore, Agerer et al. (\cite{ager96}) gave
two systems of the light elements:

for JD 2436200 -- 2445000: HJD~$(max)= 2436200.588 + 0^d.874356
\times E$;

for JD 2445000 -- 2450013: HJD~$(max)= 2449170.518 + 0^d.8744654
\times E$.

\begin{figure}
   \centering
   \includegraphics[width=11cm, angle=0]{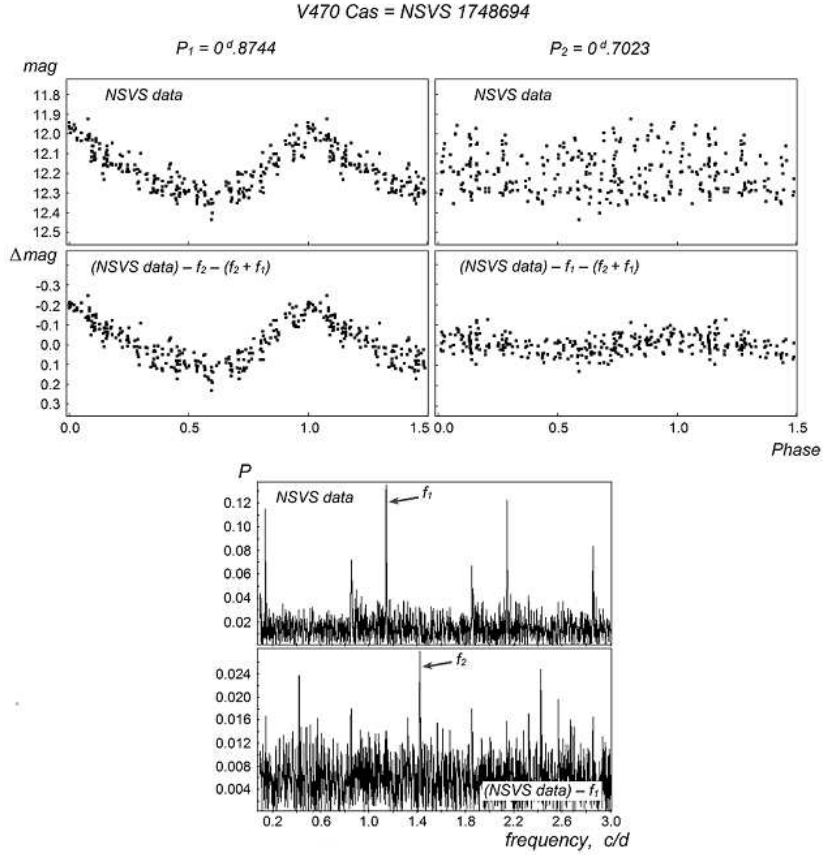}
   \caption{Light curves and power spectra of V470 Cas according to
NSVS data. The light curves in upper panels: raw data; those in
the lower panels: the folded light curves with the other
oscillation pre-whitened. Under the light curves, we present power
spectra, for the raw data and after subtraction of the first
overtone oscillations.}
   \label{Fig1}
   \end{figure}

\newpage

\subsubsection{Analysis of NSVS and SWASP data}

We suspected double-mode variability of V470~Cas from NSVS and
1SWASP data. The second frequency was detected sufficiently
reliably in the NSVS data (for this analysis, we excluded data
with the errors $err>0^m.05$); its detection in the 1SWASP data is
much less certain. The light curves and power spectra are
displayed in Figs.~2 and~3.

Our classification of V470 Cas is confirmed with its small
galactic latitude ($b = -5^{\circ}.9$) and color indices $J-K =
0.48$ (2MASS), $B-V = 0.94$ (Tycho2), and $B-V = 0.84$ (APASS),
typical for Cepheids.

   \begin{figure}
   \centering
   \includegraphics[width=11cm, angle=0]{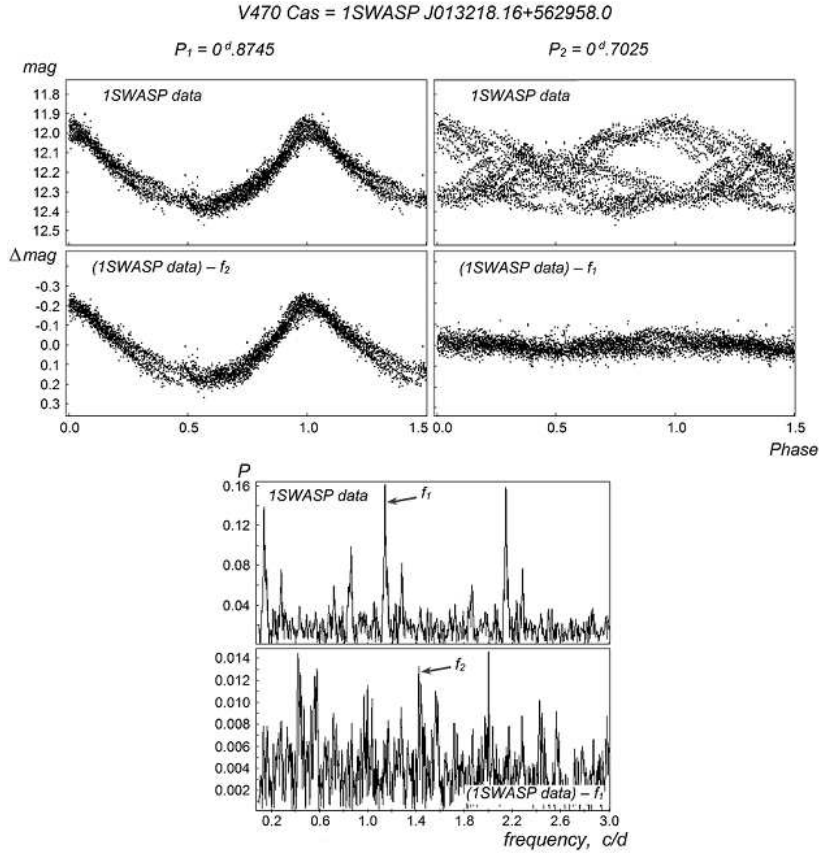}
   \caption{Light curves and power spectra of V470 Cas according to
1SWASP data.}
   \label{Fig1}
   \end{figure}

\subsubsection{CCD observations}

Our CCD observations completely confirmed the double-mode nature
of V470~Cas: this star is a double-mode Cepheid, pulsating in the
first and second overtone modes. The period ratio $P_2 /
P_1=0.8029$ is typical of the double-mode 1O/2O variables.

\newpage

The light elements of the two pulsations for all data sets are
presented in Table~2. Semi-amplitudes of the individual
oscillations and variability ranges in individual bands are
collected in Table~3. Besides the first and second overtone
frequencies, we detected one interaction frequency, $f_2+f_1$, of
V470~Cas in our CCD data.

CCD light curves of V470~Cas in the $B$, $V$, and $R$ bands are
displayed in Figs.~4 and~5. The power spectra according to CCD
observations are displayed in Fig.~6. The structure of the power
spectra leaves no doubt that $f_2$ is a real frequency.

\begin{table}
\begin{center}
\caption[]{ Light elements of V470 Cas.}\label{Tab:publ-works}

 \begin{tabular}{lllllc}
  \hline\noalign{\smallskip}
Data & $P_1$, days & Epoch$_1$, HJD & $P_2$, days & Epoch$_2$, HJD & $P_{2+1}$, days                    \\
  \hline\noalign{\smallskip}
NSVS & 0.8744 & 2451510.732 & 0.7023 & 2451510.914 & --                    \\
1SWASP & 0.8745 & 2454390.438 & 0.7025 & 2454390.425 & --                    \\
CCD & 0. 87454 & 2456789.16 & 0. 70217 & 2456789.67 & 0.389467                    \\
  \noalign{\smallskip}\hline
\end{tabular}
\end{center}
\end{table}

   \begin{figure}
   \centering
   \includegraphics[width=9cm, angle=0]{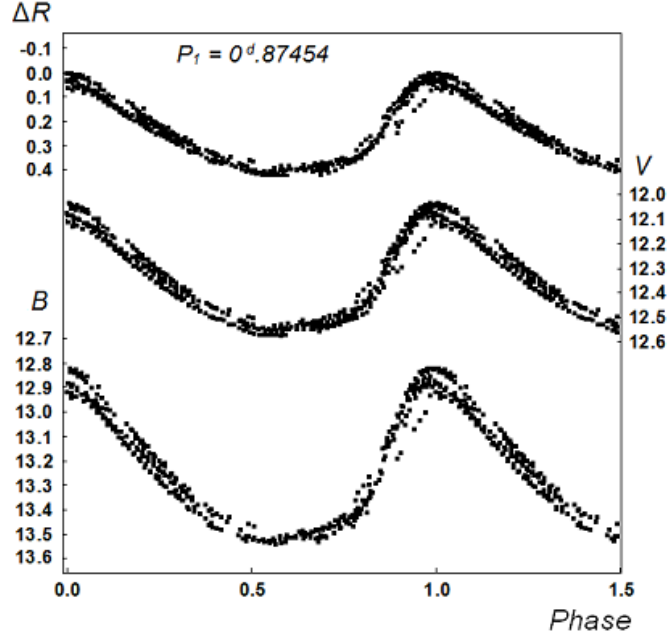}
   \caption{CCD observations: light curves in $B$, $V$ and $R$ bands for V470~Cas;
   raw data for the 1O period}
   \label{Fig1}
   \end{figure}

\newpage

\begin{table}
\begin{center}
\caption[]{ Semi-amplitudes and variability ranges of V470
Cas.}\label{Tab:publ-works}

 \begin{tabular}{llllc}
  \hline\noalign{\smallskip}
Band & $A_1$ & $A_2$ & $A_{2+1}$ & mag                    \\
  \hline\noalign{\smallskip}
NSVS (R) & 0.137 & 0.026 & -- & 11.95 -- 12.40                    \\
1SWASP & 0.189 & 0.015 & -- & 11.92 -- 12.42                    \\
$B$ & 0.3327 & 0.0364 & 0.0176 & 12.82 -- 13.54                    \\
$V$ & 0.2447 & 0.0252 & 0.0116 & 12.03 -- 12.58                    \\
$R$ & 0.1910 & 0.0185 & 0.0092 & 0.42                    \\

  \noalign{\smallskip}\hline
\end{tabular}
\end{center}
\end{table}

   \begin{figure}
   \centering
   \includegraphics[width=9cm, angle=0]{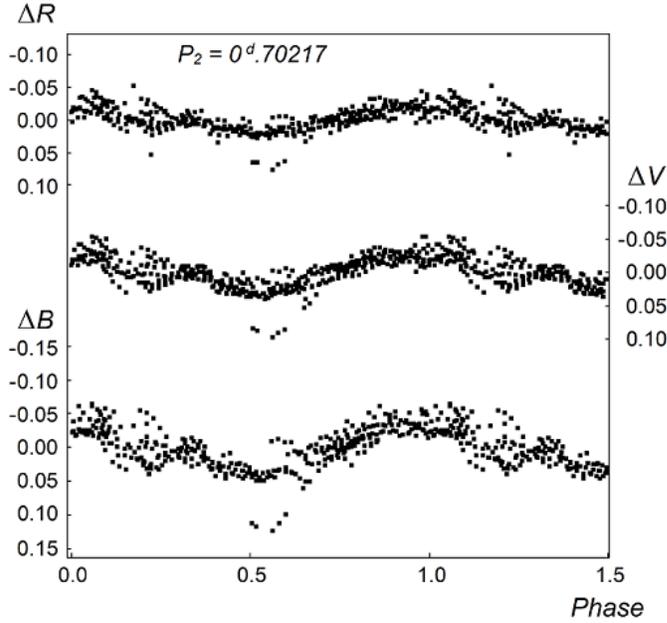}
   \caption{CCD observations: light curves in $B$, $V$, and $R$ bands for V470~Cas;
   the folded light curves for the 2O period with the other oscillation pre-whitened.}
   \label{Fig1}
   \end{figure}

\subsubsection{Period variations}

Period variations of the first overtone oscillation ($P_1$) can be
represented by an $O-C$ diagram, see Fig.~7. The parabolic shape
of this diagram is typical of secular period variations of
classical Cepheids. Figure~6 is based on linear light elements for
the middle of the time interval:

HJD~$(max) = 2446000.633 + 0^d.874434 \times E$.

Before JD 2449000, all points are times of high brightness
according to photographic photometry; after JD 2449000, all points
are CCD maxima. We used the data from Agerer et al.
(\cite{ager96}), Agerer \& Huebscher (\cite{ager02}), Huebscher
(\cite{hueb05}), Huebscher et al. (\cite{hueb05et}), Huebscher et
al. (\cite{hueb06}), Huebscher et al. (\cite{hueb09}), Huebscher
et al. (\cite{hueb10}), Huebscher \& Lehmann (\cite{hueb12}), and
data from our study.

\newpage

   \begin{figure}
   \centering
   \includegraphics[width=9cm, angle=0]{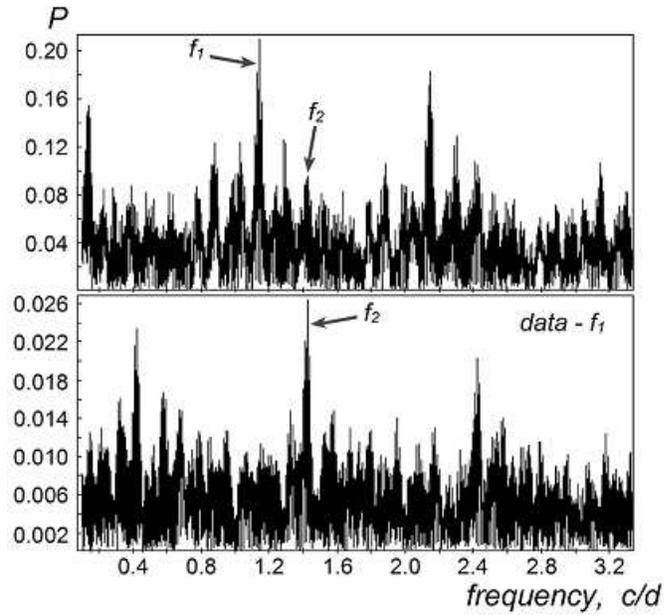}
   \caption{Power spectra of V470 Cas for the frequencies $f_1$ and
$f_2$ according to CCD observations in the $V$ band. Upper panel:
raw data; lower panel: the first overtone oscillation
pre-whitened.}
   \label{Fig1}
   \end{figure}

   \begin{figure}
   \centering
   \includegraphics[width=9cm, angle=0]{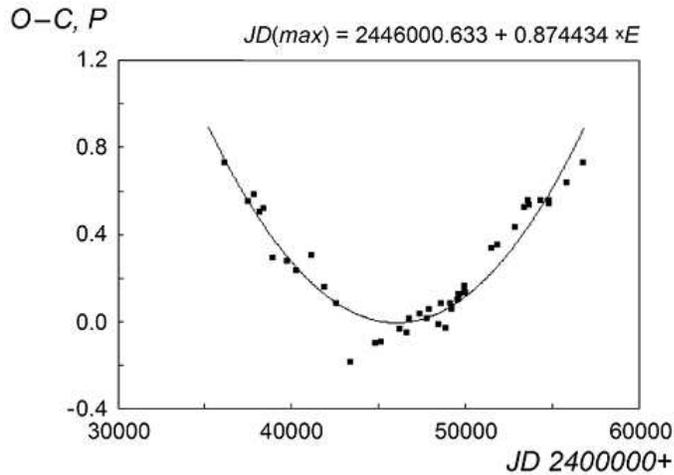}
   \caption{$O-C$ diagram for the period of the first overtone
oscillation of V470~Cas.}
   \label{Fig1}
   \end{figure}

\subsection{GSC 2901-00089}

\subsubsection{Earlier studies}

The variability of GSC 2901-00089 ($\alpha = 04^h 45^m 23^s.89,
\delta = +42^{\circ} 55^{\prime} 20^{\prime\prime}.0$, J2000) was
reported by Hoffman et al. (\cite{hoff09}) from ROTSE-I/NSVS data
(NSVS~4346946). The variable was classified as an RR Lyrae star
with the period of 0.53391 days. 

\newpage

Later, we classified GSC
2901-00089 as a 1O/2O double-mode Cepheid (Khruslov~\cite{khru13})
using all available observations from the ROTSE-I/NSVS and
SuperWASP online public archives. In the cited paper, we presented
the preliminary results. The paper contained only the two periods
and the new classification, CEP(B) variability type in the GCVS
classifications system (Samus et al.~\cite{samu15}). Now we have
re-analysed NSVS and SuperWASP data and our CCD observations,
confirmed the 1O/2O double periodicity of GSC 2901-00089, and
improved its light elements.

   \begin{figure}
   \centering
   \includegraphics[width=11cm, angle=0]{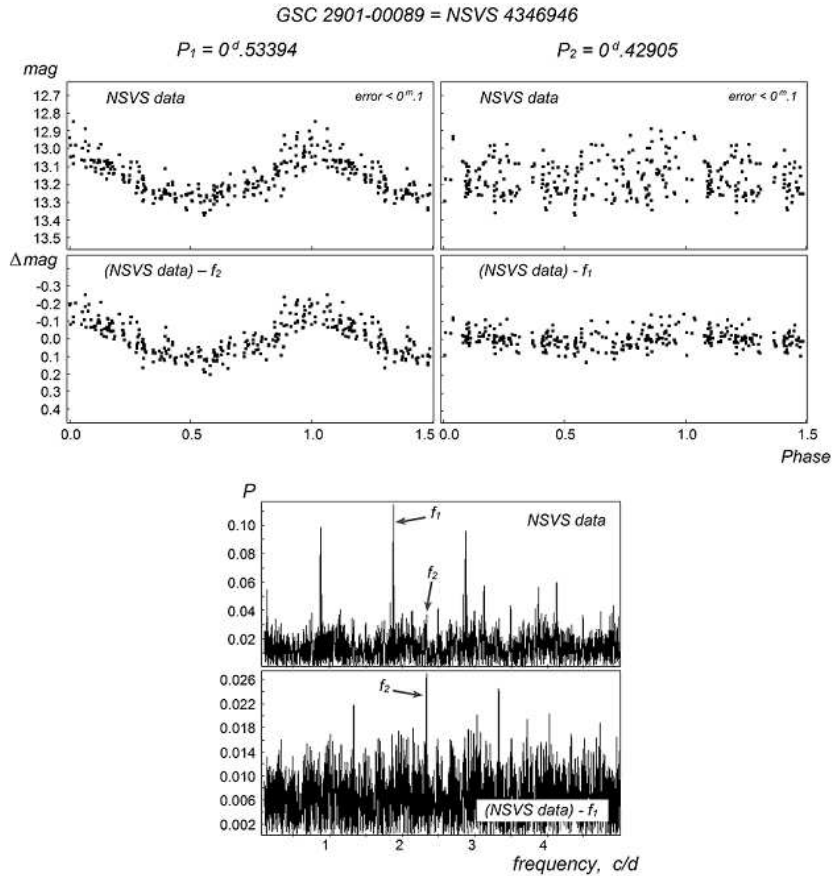}
   \caption{Light curves and power spectra of GSC 2901-00089 according
to NSVS data. The light curves in upper panels: raw data; those in
the lower panels: the folded light curves with the other
oscillation pre-whitened. Under the light curves, we present power
spectra, for the raw data and after subtraction of the first
overtone oscillations.}
   \label{Fig1}
   \end{figure}

\newpage

\subsubsection{Analysis of NSVS and SWASP data}

The light curves and power spectra from NSVS and SuperWASP data
are displayed in Figs.~8 and 9. For the analysis of NSVS data, we
excluded data with the errors $err > 0^m.1$; for the analysis of
SWASP data, we excluded data with $err > 0^m.05$. The ROTSE data
with photometric correction flags (usually rejected) were kept for
the analysis.

Our classification is confirmed by the small galactic latitude $b
= -1^{\circ}.7$ and by the color indices $J - K = 0.54$ (2MASS),
$B - V = 1.05$ (APASS).

\subsubsection{CCD observations}

Our CCD observations completely confirmed the double-mode nature
of GSC 2901-00089: this star is a 1O/2O double-mode Cepheid. The
period ratio $P_2/P_1 = 0.8036$ is typical of variables of this
type. Besides the first and second overtone frequencies, we
detected two interaction frequencies of GSC 2901-00089, $f_2 +
f_1$ and $f_2 - f_1$, in our CCD data.

   \begin{figure}
   \centering
   \includegraphics[width=11cm, angle=0]{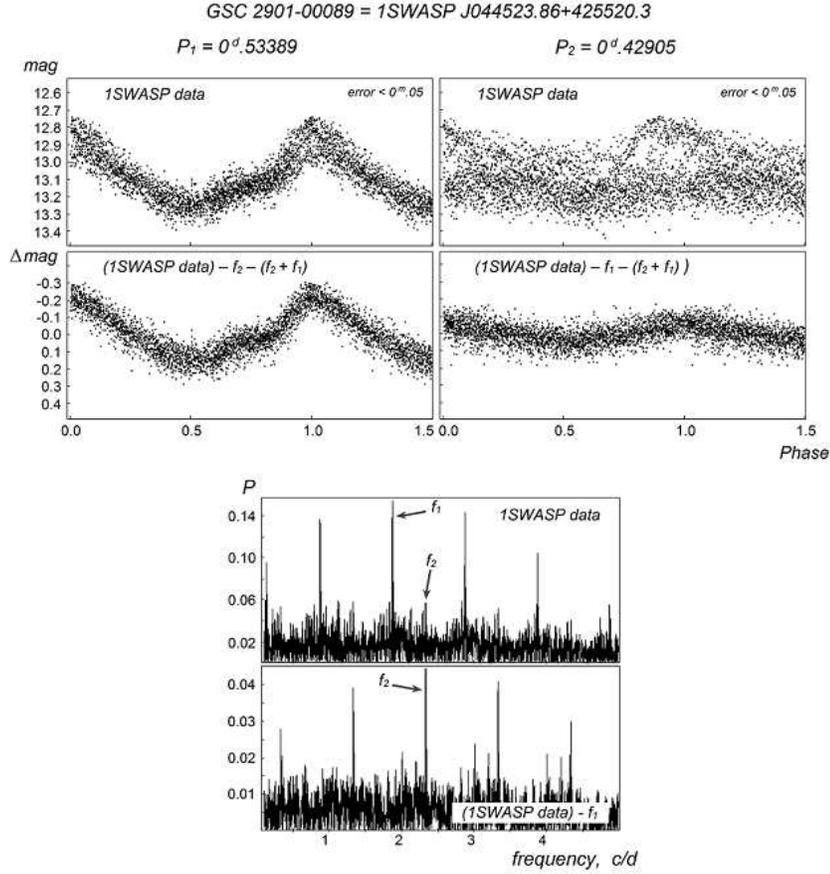}
   \caption{Light curves and power spectra of GSC 2901-00089 according
to 1SWASP data.}
   \label{Fig1}
   \end{figure}

\newpage

The light elements of the two pulsations for all data sets are
presented in Table~4:  the first-overtone period $P_1$ and epoch,
the second-overtone period $P_2$ and epoch, periods for the
frequencies $f_2 + f_1$ and $f_2 - f_1$. The periods and epochs
are given for all individual data sets. Semi-amplitudes of the
individual oscillations and the variability ranges in different
bands are collected in Table~5. For the $R$ band, we give the full
variability amplitude (peak to peak).

The CCD light curves in the $B$, $V$, and $R$ bands for
GSC~2901-00089 are displayed in Figs.~10 and 11. The power spectra
according to CCD observations are shown in Fig.~12. The structure
of the power spectra leaves no doubt that $f_2$ is a real
frequency.

\begin{table}
\begin{center}
\caption[]{ Light elements of GSC 2901-00089.
}\label{Tab:publ-works}

 \begin{tabular}{llllllc}
  \hline\noalign{\smallskip}
Data & $P_1$, days & Epoch$_1$, HJD & $P_2$, days & Epoch$_2$, HJD & $P_{2+1}$, days  & $P_{2-1}$, days                   \\
  \hline\noalign{\smallskip}
NSVS & 0.53394 & 2451450.380 & 0.42905 & 2451450.390 & -- & --                    \\
1SWASP & 0.53389 & 2453700.227 & 0.42905 & 2453700.333 & 0.23796 & --                   \\
CCD & 0.533824 & 2456650.325 & 0.428983 & 2456650.288 & 0.237848 & 2.1843                   \\

  \noalign{\smallskip}\hline
\end{tabular}
\end{center}
\end{table}

   \begin{figure}
   \centering
   \includegraphics[width=9cm, angle=0]{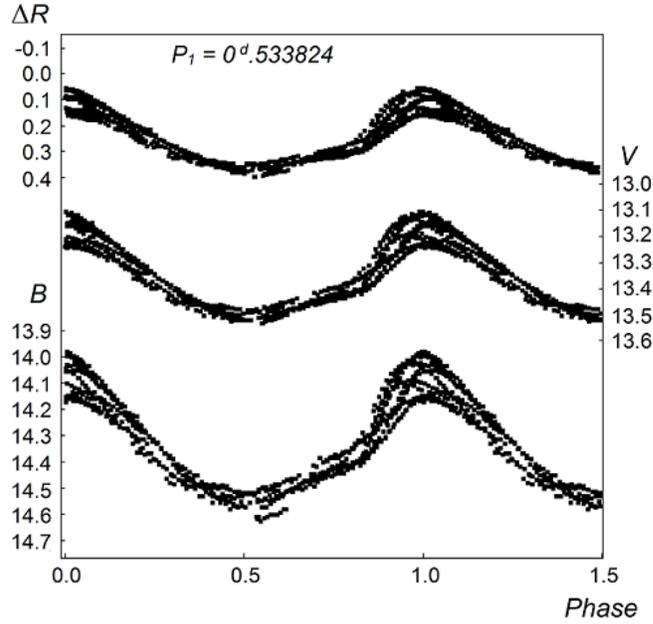}
   \caption{CCD observations: light curves in $B$, $V$, and $R$
bands for GSC 2901-00089; raw data for the 1O period.}
   \label{Fig1}
   \end{figure}

\newpage

\subsubsection{Period variations}

The periods of the first and second overtone oscillations vary
significantly, see Table~4. We can claim it beyond doubt for $P_1$
(progressive decrease). The diagram of the period variations of
GSC~2901-00089 is displayed in Fig.~13.

\section{Conclusions}
\label{sect:conclusion}

In this paper we presented a photometric study of two new
double-mode Cepheids, pulsating in the first and second overtones
mode. From our multicolor CCD observations in the Johnson's $B$,
$V$ and $R$ bands, we found the light elements of two oscillatons
and detected interaction frequencies for each of the stars. We
study period variations of the stars; variations of the first
overtone periods were reliably detected.

\begin{table}
\begin{center}
\caption[]{ Semi-amplitudes and variability ranges of GSC
2901-00089. }\label{Tab:publ-works}

 \begin{tabular}{lllllc}
  \hline\noalign{\smallskip}
Band & $A_1$ & $A_2$ & $A_{2+1}$ & $A_{2-1}$ & mag                    \\
  \hline\noalign{\smallskip}
NSVS & 0.119 & 0.029 & -- & -- & 12.9\phantom{1} -- 13.35                    \\
1SWASP & 0.179 & 0.044 & 0.017 & -- & 12.74 -- 13.37                    \\
$B$ & 0.2349 & 0.0418 & 0.0154 & 0.0115 & 13.98 -- 14.63                    \\
$V$ & 0.1626 & 0.0289 & 0.0114 & 0.0094 & 13.11 -- 13.54                    \\
$R$ & 0.1278 & 0.0209 & 0.0091 & 0.0075 & 0.34                    \\

  \noalign{\smallskip}\hline
\end{tabular}
\end{center}
\end{table}

   \begin{figure}
   \centering
   \includegraphics[width=9cm, angle=0]{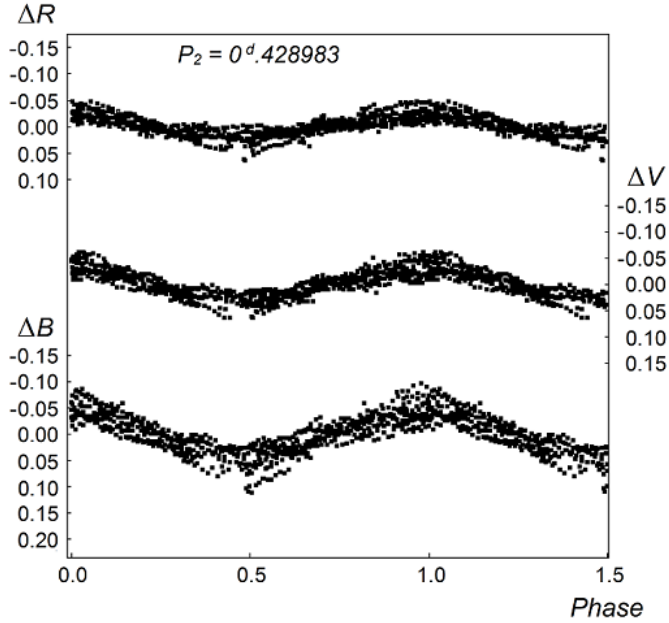}
   \caption{CCD observations: light curves in $B$, $V$, and $R$ band
for GSC 2901-00089; the folded light curves for the 2O period with
the other oscillation pre-whitened.}
   \label{Fig1}
   \end{figure}

\newpage

It is more difficult to find 1O/2O Cepheids in our Galaxy compared
to the Magellanic Clouds, and each new case if of a considerable
interest. It is already possible to use known Cepheids of this
kind, including those identified by the author, to plot the
Petersen diagram and to compare the Galactic sample of the 1O/2O
Cepheids to that of the 1O/2O Cepheids in the Magellanic Clouds.

   \begin{figure}
   \centering
   \includegraphics[width=9cm, angle=0]{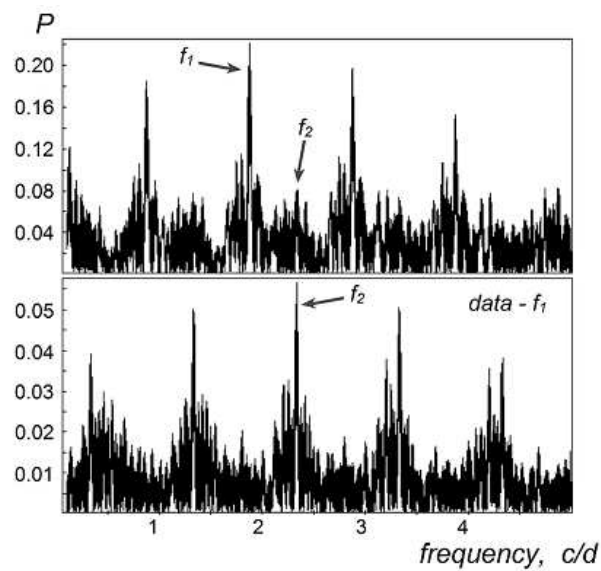}
   \caption{Power spectra of GSC 2901-00089 for the frequencies $f_1$
and $f_2$ according to CCD observations in the $B$ band. Upper
panels: raw data; lower panels: the first overtone oscillation
pre-whitened.}
   \label{Fig1}
   \end{figure}

   \begin{figure}
   \centering
   \includegraphics[width=6cm, angle=0]{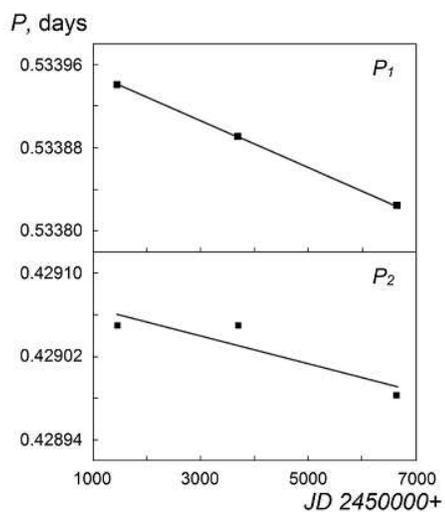}
   \caption{The period variations of GSC 2901-00089.}
   \label{Fig1}
   \end{figure}

\newpage

Figure~14 presented the Petersen diagram for Galactic 1O/2O
Cepheids compared to 1O/2O Cepheids in the LMC. Solid circles
represent known 1O/2O Galactic Cepheids; open circles are the two
1O/2O Galactic Cepheids suspected by Khruslov (2013); the solid
squares are the two Cepheids of this study; solid triangles are
three-modal Galactic Cepheids; open triangles, 1O/2O Cepheids in
the LMC (Soszynsky et al. 2008). In addition, the figure also
shows two 1O/2O stars classified not as double-mode Cepheids but
actually very similar to them: V798 and V1719 Cygni.

It appears from this diagram that the two samples differ
considerably, probably because of chemical-composition differences
between the galaxies.

   \begin{figure}
   \centering
   \includegraphics[width=\textwidth, angle=0]{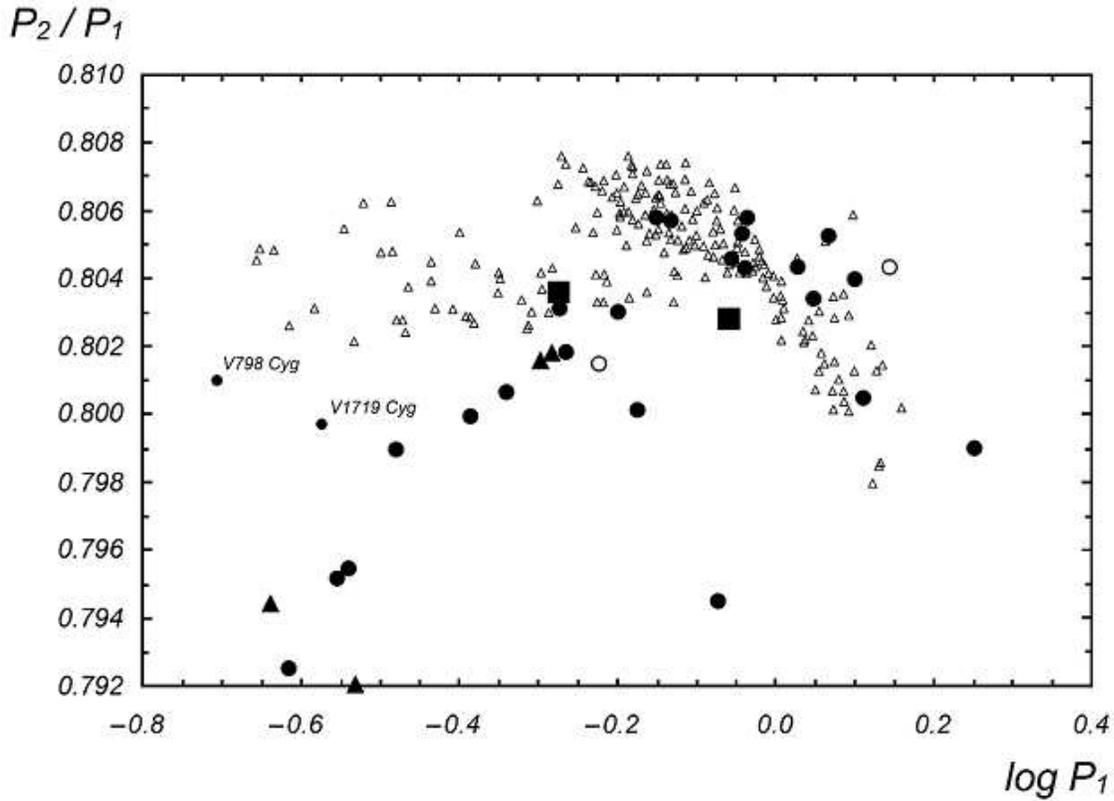}
   \caption{The Petersen diagram for the Galactic 1O/2O Cepheids
   compared to the 1O/2O Cepheids in the LMC. }
   \label{Fig1}
   \end{figure}

\newpage

\begin{acknowledgements}
The authors are grateful to Dr. V. P. Goranskij for providing
light-curve analysis software. Thanks are due to Drs. S.V. Antipin
and N.N. Samus for helpful discussions and to Dr. K.V. Sokolovsky
for his advice concerning data retrieving. We wish to thank M. A.
Krugov, N. V. Lichkanovsky, I. V. Rudakov, I. V. Reva, R. I.
Kokumbaeva and W. Mundrzyjewski for their assistance during the
observations. This study was supported by the Russian Foundation
for Basic Research (grant 13-02-00664), the Basic Research Program
P-7 of the Presidium of Russian Academy of Sciences, and the
program "Studies of Physical Phenomena in Star-forming Regions and
Nuclear Zones of Active Galaxies" of the Ministry of Education and
Science (Republic of Kazakhstan).
\end{acknowledgements}

\label{lastpage}

\end{document}